\documentclass[a4paper,12pt]{article}
\usepackage[dvips]{graphicx}
\DeclareGraphicsExtensions{.pdf,.jpg.eps}

\usepackage{amsfonts,amsmath,amsthm,amssymb}
\usepackage{mathrsfs}
\usepackage[onehalfspacing]{setspace}
\author{L. Ort\'{i}z\footnote{leonardo.ortiz@uniofyorkspace.net}}

\title{Hawking temperature in the eternal BTZ black hole: an example of
Holography in AdS spacetime}

\begin{document}

\maketitle

\begin{center}



Department of Mathematics\\
The University of York\\
York YO10 5DD, U. K.\\\vspace{0.4cm}

Department of Physics\\
University of Guanajuato\\
Leon Guanajuato 37150, Mexico\\\vspace{0.4cm}

\normalsize{\textbf{Abstract}}\\\end{center} \small{We review the
relation between AdS spacetime in 1+2 dimensions and the BTZ black
hole. Later we show that a ground state in AdS spacetime becomes a
thermal state in the BTZ black hole. We show that this is true in
the bulk and in the boundary of AdS spacetime. The existence of
this thermal state is tantamount to say that the Unruh effect
exists in AdS spacetime and becomes the Hawking effect for an
eternal BTZ black hole. In order to make this we use the
correspondence introduced in Algebraic Holography between algebras
of quasi-local observables associated to wedges and double cones
regions in the bulk of AdS spacetime and its conformal boundary
respectively. Also we give the real scalar quantum field as a
concrete heuristic realization of this formalism.}

\section{Introduction}

\normalsize{In} recent years there has been great interest in the
so-called AdS/CFT correspondence. It seems fair to say that the
majority of works on this topic belong to the string theory
framework. However, in the context of Quantum Field Theory (QFT),
also some approaches to the AdS/CFT correspondence have been
proposed. More precisely, in the context of Algebraic Quantum
Field Theory (AQFT) there appeared Algebraic Holography (AH)
\cite{khReh00}. Algebraic Holography relates a covariant quantum
field theory in the bulk and a conformally covariant quantum field
theory on the conformal boundary\footnote{From here on instead of
writing conformal boundary of AdS spacetime we will just write
boundary of AdS spacetime unless a confusion can arise.} of AdS
spacetime. In this sense AH gives an AdS/CFT correspondence too.
Also there has appeared the boundary-limit holography
\cite{mBerjBrosuMosrSch00} where it has been constructed a
correspondence between $n$-point functions of a covariant quantum
field theory in the bulk and a conformally covariant quantum field
theory in the boundary of AdS spacetime. More recently, partly
motivated for these works, there appeared Pre-Holography
\cite{pLarbKay08}, which studied some aspects of the
correspondence between field theories in the bulk and in the
boundary of AdS spacetime by using the symplectic structure
associated to the phase space of the Klein-Gordon operator. When
we take into account these works, it is clear that even at the
level of QFT there are interesting aspects of the AdS/CFT
correspondence which deserve to be studied. Amongst these aspects
of the AdS/CFT in QFT\footnote{By AdS/CFT in QFT we mean AH, the
boundary-limit holography and Pre-Holography since all them fit in
the QFT framework.} which has been studied so far it is, for
example, how the global ground state in the bulk of AdS spacetime
maps to a state in its boundary \cite{pLarbKay08}.

In QFT besides ground states the equilibrium thermal states are
outstanding elements in the theory. So, after studying the mapping
of a ground state it is very natural to ask how an equilibrium
thermal state maps from the bulk to the boundary of AdS spacetime.
This issue is more appealing when one knows that models of black
holes can be obtained from AdS spacetime. In particular it is
known that in three dimensions there exists a solution to the
Einstein's field equations which can be considered as a model of a
black hole, the BTZ black hole (BTZbh) \cite{BHTZ93}. This
solution can be obtained directly from the Einstein's field
equations or by making identifications in a proper subset of AdS
spacetime \cite{BHTZ93}.

The purpose of this work is to study how a ground state in AdS
spacetime in three dimensions is related to an equilibrium thermal
state in the BTZbh and how it maps to its boundary. We address
this issue by using the abstract setting of Algebraic Holography
and by considering a quantum real scalar field. One conclusion we
obtain from this investigation is that the Unruh effect takes
place in the boundary and in the bulk of AdS spacetime in 1+2
dimensions, and that after a quotient procedure this effect
becomes the Hawking effect for the eternal BTZ black hole. This
work also will give details of calculations that will be used in
the complementary work \cite{bKaylOrt11}.

In our study the symmetries of AdS spacetime are fundamental. The
principal fact is that the group $SO_{0}(2,2)$ acts in AdS
spacetime and in its conformal boundary. In AdS spacetime it acts
as the isometry group and in its conformal boundary as the global
conformal group. This is fundamental in making AH possible. Also,
although indirectly, the Tomita-Takesaki and Wichmann-Bisogano
theorems play a fundamental r\^{o}le. These mathematical aspects
are relevant when addressing our problem in the abstract setting.
When we give a concrete heuristic example of this formalism we use
the procedure given in \cite{mBerjBrosuMosrSch00} to obtain a two
point function in the boundary of AdS spacetime from a two point
function in its bulk.

This work is organized as follows: in section 2 we introduce some
aspects of AdS spacetime in 1+2 dimensions relevant for this work.
In section 3 we give the generalities of the BTZbh. In section 4
we study the relation of the BTZbh and the boundary of AdS
spacetime. In section 5 we define wedge regions in AdS spacetime
and show that the exterior of the BTZbh is a wedge regions. Also
we proof that the subgroup of AdS group which leaves invariant the
exterior of the BTZbh corresponds to the Lorentz boost which
leaves invariant the Rindler wedge of the $2$-dimensional
Minkowski spacetime at infinity of AdS spacetime. In section 6 we
proof that an equilibrium thermal state in AdS spacetime maps to
an equilibrium thermal state in the BTZbh. This state is invariant
under translations of BTZ time. Also we give as a heuristic
realization of this formalism the real quantum scalar field. In
appendix A we construct the finite transformations of the global
conformal group in two dimensions.

\section{AdS spacetime in 1+2 dimensions}

\normalsize{In} 1+2 dimensions AdS spacetime can be introduced as follows: Let
us consider $\mathbb{R}^{4}$ endowed with metric
\begin{equation}\label{E:1}
ds^{2}=-du^{2}-dv^{2}+dx^{2}+dy^{2}.
\end{equation}
We denote the resulting semi-Euclidian space by
$\mathbb{R}^{2,2}$. AdS spacetime in 1+2 dimensions can be
identified with the hypersurface in $\mathbb{R}^{2,2}$ defined by
\begin{equation}\label{E:104}
-u^{2}-v^{2}+x^{2}+y^{2}=-\ell^{2}
\end{equation}
for fixed $\ell$ and with metric induced by the pull back of
(\ref{E:1}) to (\ref{E:104}) under the inclusion map.

At infinity the limit of (\ref{E:104}) is \cite{eWitt98}
\begin{equation}\label{E:104aaa}
-u^{2}-v^{2}+x^{2}+y^{2}=0,
\end{equation}
which we call the null cone and denote by $\mathcal{C}^{4}$. The projective cone\footnote{The projective cone is obtained from the null cone by identifying a ray in the null cone with a point. This is why we can introduce the coordinates (\ref{E:104aaaa}) in this projective cone.} obtained from the null cone is the compactification
of a two dimensional Minkowski spacetime. In this Minkoswski spacetime
we can introduce coordinates \cite{rHaag96} p. 15
\begin{equation}\label{E:104aaaa}
\xi^{1}=\frac{u}{v+x}\hspace{1cm}\xi^{2}=\frac{y}{v+x}
\end{equation}
and a metric $\eta_{\mu\nu}=\textrm{diag}(-1,1)$. These
coordinates do not cover all the manifold, points at infinity are
left out. Both manifolds, (\ref{E:104}) and (\ref{E:104aaa}), are
invariant under $X\leftrightarrow -X$, hence AdS spacetime modulo
this identification has a compactified Minkowski spacetime at
infinity. From (\ref{E:104aaa}) we see that the topology of
$\mathcal{C}^{4}$ is the topology of
$\mathbb{S}^{1}\times\mathbb{S}^{1}/{\pm I}$.

The expressions (\ref{E:104}) and (\ref{E:104aaa}) both
are invariant under the orthogonal group $O(2,2)$. In particular they are
invariant under its connected component, namely $SO_{0}(2,2)$.
However the action of this group has a different meaning when
acting on the hypersurfaces defined by these expressions.
In the first case it acts as rotations of $\mathbb{R}^{2,2}$ which
preserve (\ref{E:104}), and we call it AdS group; whereas by
preserving (\ref{E:104aaa}) it acts on (\ref{E:104aaaa}) as the
global conformal group in $2$-dimensional Minkowski spacetime.



Now let us introduce global and Poincar\'{e} charts.

\subsection{Global coordinates}

Global coordinates $(\lambda,\rho,\theta)$ can be defined by
\begin{eqnarray}\label{E:105}
v=\ell\sec\rho\cos\lambda\hspace{1cm}u=\ell\sec\rho\sin\lambda\nonumber\\
x=\ell\tan\rho\cos\theta\hspace{1cm}y=\ell\tan\rho\sin\theta,
\end{eqnarray}
where
$(\lambda,\rho,\theta)\in[-\pi,\pi)\times[0,\pi/2)\times[-\pi,\pi)$.
In these coordinates the metric is
\begin{equation}\label{E:106aa}
ds^{2}=\ell^{2}\sec^{2}\rho\left(-d\lambda^{2}+d\rho^{2}+\sin^{2}\rho
d\theta^{2}\right).
\end{equation}
From (\ref{E:105}) we see that $\rho\rightarrow\pi/2$ corresponds
to infinity. The metric (\ref{E:106aa}) is not defined on this
point, however we can define an usually called unphysical metric
as $d\tilde{s}^{2}=\Omega^{2}ds^{2}$ with
$\Omega=\frac{1}{\ell}\cos\rho$ and get
$d\tilde{s}^{2}=-d\lambda^{2}+d\rho^{2}+\sin^{2}\rho d\theta^{2}$.
This metric is well defined for $\rho\in[0,\pi/2]$. When
constructing a Penrose diagram for AdS spacetime this is the
metric most commonly used \cite{sHawgEll84}. This is the metric of the Einstein universe,
but it cover just half of it since $\rho\in[0,\pi/2]$. Using this
conformal mapping we can attach a boundary to AdS spacetime. This
boundary is given by $\rho=\pi/2$ and its metric is
\begin{equation}\label{E:106b}
d\tilde{s}^{2}_{b}=-d\lambda^{2}+ d\theta^{2}.
\end{equation}
The boundary is
$\mathbb{S}^{1}\times\mathbb{S}^{1}$ if we make the identification of $-\pi$ and $\pi$ in the domain of $\lambda$ and $\theta$. We can implement $X\leftrightarrow -X$ by antipodal identification
in these two circles. In order to avoid close
timelike curves it is customary to work with the covering space of
AdS spacetime (CAdS), i.e., by letting $\lambda$ to vary on
$\mathbb{R}$, and then the boundary of CAdS is an infinite long
cylinder $\mathbb{R}\times\mathbb{S}^{1}$. It is again an Einstein
universe but in two dimensions.

\subsection{Poincar\'{e} coordinates}

Poincar\'{e} coordinates $(T,k,z)$ are given by \cite{caBallrfBra07}
\begin{eqnarray}\label{E:107}
v=\frac{1}{2z}\left(z^{2}+\ell^{2}+k^{2}-T^{2}\right)\hspace{1cm}u=\frac{\ell T}{z}\nonumber\\
x=\frac{1}{2z}\left(\ell^{2}-z^{2}+T^{2}-k^{2}\right)\hspace{1cm}y=\frac{\ell
k}{z}.
\end{eqnarray}
In these coordinates the metric is
\begin{equation}\label{E:108}
ds^{2}=\frac{\ell^{2}}{z^{2}}\left(-dT^{2}+dk^{2}+dz^{2}\right)
\end{equation}
where
$(T,k,z)\in(-\infty,\infty)\times(-\infty,\infty)\times(0,\infty)$. In these coordinates, $X\leftrightarrow -X$ can be implemented by $z\leftrightarrow-z$.


\section{The BTZ black hole}

\normalsize{It} is well known there exists a solution to the
Einstein's field equations with negative cosmological constant in
three dimensions which can be considered as a model of a black
hole \cite{BHTZ93}, better known as the BTZ black hole (BTZbh).
The metric of this spacetime is
\begin{equation}\label{E:109}
ds^{2}=-f^{2}dt^{2}+f^{-2}dr^{2}+r^{2}(d\phi+N^{\phi}dt)^{2},
\end{equation}
where
$f^{2}=\left(-M+\frac{r^{2}}{\ell^{2}}+\frac{J^{2}}{4r^{2}}\right)$
and $N^{\phi}=-\frac{J}{2r^{2}}$
with $|J|\leq M\ell$. We call
$(t,r,\phi)\in(-\infty,\infty)\times(0,\infty)\times[0,2\pi)$ BTZ
coordinates. This metric is asymptotically AdS spacetime since
when $r\rightarrow\infty$
\begin{equation}\label{E:b3}
ds^{2}\sim r^{2}\left(-dt^{2}+d\phi^{2}\right).
\end{equation}
Hence the BTZbh is asymptotically
$\mathbb{R}\times\mathbb{S}^{1}$, an infinite long cylinder. The horizons
are given by
\begin{equation}\label{E:111}
r_{\pm}^{2}=\frac{M\ell^{2}}{2}\left(1\pm\left(1-\left(\frac{J}{Ml}\right)^{2}\right)^{1/2}\right)
\end{equation}
The BTZ metric can be obtained directly from the
Einstein's field equations by imposing time and axial symmetry
\cite{BHTZ93}. Also, it can be obtained by identifying points in AdS
spacetime along the orbits of an appropriate Killing vector of the
AdS spacetime \cite{BHTZ93}. When expressed in BTZ
coordinates this Killing vector turns out to be $\partial_{\phi}$.

For the purposes of this work we are interested in the exterior
region of the black hole, $r>r_{+}$. This region can be
parameterized by \cite{BHTZ93}
\begin{eqnarray}\label{E:112}
u=\sqrt{B(r)}\sinh\tilde{t}\left(t,\phi\right)\hspace{1cm}v=\sqrt{A(r)}\cosh\tilde{\phi}\left(t,\phi\right)\nonumber\\
y=\sqrt{B(r)}\cosh\tilde{t}\left(t,\phi\right)\hspace{1cm}x=\sqrt{A(r)}\sinh\tilde{\phi}\left(t,\phi\right).
\end{eqnarray}
where
\begin{equation}\label{E:113}
A(r)=\ell^{2}\left(\frac{r^{2}-r_{-}^{2}}{r_{+}^{2}-r_{-}^{2}}\right),\hspace{1cm}B(r)=\ell^{2}\left(\frac{r^{2}-r_{+}^{2}}{r_{+}^{2}-r_{-}^{2}}\right)
\end{equation}
and
\begin{equation}\label{E:113a}
\tilde{t}=\frac{1}{\ell}\left(r_{+}t/\ell-r_{-}\phi\right),\hspace{1cm}\tilde{\phi}=\frac{1}{\ell}\left(r_{+}\phi-r_{-}t/\ell\right).
\end{equation}
In this parametrization the coordinate $\phi$ must be
$2\pi$-periodic in order to obtain the BTZ metric.


An important quantity in our study of thermal states in BTZbh is the surface gravity, $\kappa$. This turns out to be \cite{sCar98}
$\kappa=\frac{r_{+}^{2}-r_{-}^{2}}{\ell^{2}r_{+}}$.

\begin{figure}
\includegraphics[scale=0.4]{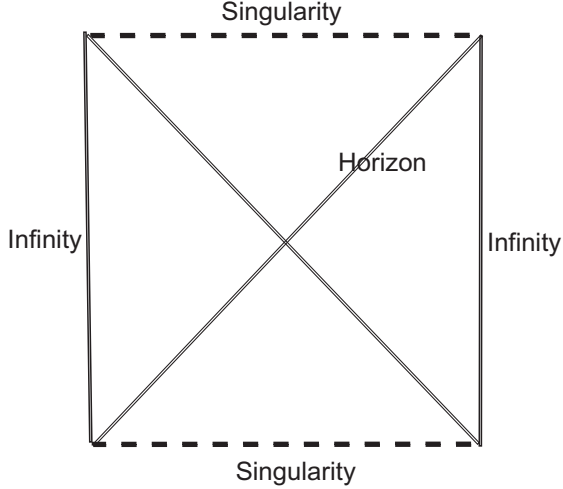}
\caption{\label{F:penrose}\footnotesize{Penrose diagram for the
non-rotating BTZ black hole.}}
\end{figure}

For further reference we give in figure \ref{F:penrose} the
Penrose diagram for the non-rotating BTZbh.

\section{The BTZ black hole and its relation to the boundary of AdS spacetime}

We mentioned before that the BTZbh can be obtained by a quotient
procedure from AdS spacetime. This quotient is made by a discrete
subgroup of $SO_{0}(2,2)$\footnote{This subgroup is a subgroup of period 2$\pi$ of the continuous group generated by the appropriate Killing vector of AdS spacetime.}. Because we want the resulting spacetime
not to have closed timelike curves, it is required the Killing
vector which generates this subgroup to be spacelike. This
criterion turns out to be not just necessary but also sufficient
\cite{BHTZ93}. First let us restrict ourselves to the non-rotating
case. In this case, when expressed in embedding coordinates this
generator turns out to be
\begin{equation}\label{E:12}
\partial_{\phi}=\frac{r_{+}}{\ell}\left(x\partial_{v}+v\partial_{x}\right).
\end{equation}
We are interested in studying quantum field theory in the boundary
of AdS spacetime and consequently of the BTZbh. So it is useful to
find out which regions of the boundary of AdS spacetime correspond
to the covering space of the BTZbh. Because the maximally extended BTZbh have two exterior regions
analogously to Schwarzschild spacetime, there will be two regions
in the boundary which cover the maximally extended BTZbh. If we
want to know explicitly these two regions, we can express
$\partial_{\phi}$ in global coordinates, take the limit
$\rho\rightarrow\pi/2$, and impose on it the condition of being
spacelike in the metric (\ref{E:106b}).

Using (\ref{E:105}) and (\ref{E:12}), in global coordinates
\begin{equation}\label{E:14}
\frac{\ell}{r_{+}}\partial_{\phi}=-\sin\lambda\sin\rho\cos\theta\partial_{\lambda}+\cos\lambda\cos\rho\cos\theta\partial_{\rho}-\frac{\cos\lambda\sin\theta}{\sin\rho}\partial_{\theta}.
\end{equation}
On the boundary
$J_{\phi}\equiv\frac{\ell}{r_{+}}\partial_{\phi}|_{\rho=\pi/2}=-\sin\lambda\cos\theta\partial_{\lambda}-\cos\lambda\sin\theta\partial_{\theta}$.
Its norm is given by
$||J_{\phi}||^{2}=-\sin^{2}\lambda\cos^{2}\theta+\cos^{2}\lambda\sin^{2}\theta=-\sin
u\sin v$,
where we have introduced null coordinates
$u=\lambda-\theta$ and $v=\lambda+\theta$.
Clearly this vector can be timelike, spacelike or null. The
regions where it is null are given by
$u,v=n\pi$ with $n=0,\pm 1,\pm 2 ...$
Hence the covering space of the exterior of the BTZbh is the
region inside the lines defined by,
\begin{equation}\label{E:19}
u=\pi,0,-\pi\hspace{1cm}v=\pi,0,-\pi,
\end{equation}
see figure 2.
\begin{figure}\centering
\includegraphics[scale=0.4]{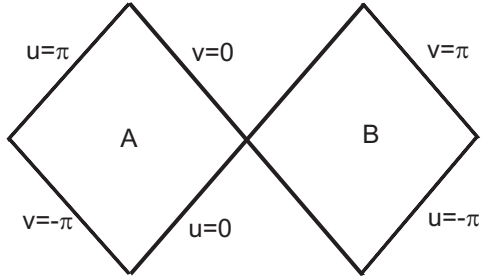}
\caption{\label{F:btzboundary}\footnotesize{This figure represents
the covering space of the BTZbh in the boundary of AdS spacetime
in 1+2 dimensions.}}
\end{figure}
This figure has been given in \cite{sAminiBensHol99} and
\cite{dBrill} too.

On the boundary the BTZ coordinates are $(t,\phi)$. We have found
already the generator of translations in $\phi$, let us see what
the translations in time is. In embedding coordinates
\begin{equation}\label{E:20}
\partial_{t}=\frac{r_{+}}{\ell^{2}}\left(y\partial_{u}+u\partial_{y}\right).
\end{equation}
Using (\ref{E:105}) and (\ref{E:20}), in global coordinates
\begin{equation}\label{E:21}
\frac{\ell^{2}}{r_{+}}\partial_{t}=\sin\lambda\cos\rho\sin\theta\partial_{\rho}+\cos\lambda\sin\rho\sin\theta\partial_{\lambda}+\frac{\sin\lambda\cos\theta}{\sin\rho}\partial_{\theta}.
\end{equation}
On the boundary
$J_{t}\equiv\frac{\ell^{2}}{r_{+}}\partial_{t}|_{\rho=\pi/2}=\cos\lambda\sin\theta\partial_{\lambda}+\sin\lambda\cos\theta\partial_{\theta}$.
The norm of $J_{t}$ is given by
$||J_{t}||^{2}=-\cos^{2}\lambda\sin^{2}\theta+\sin^{2}\lambda\cos^{2}\theta=\sin
u\sin v$.
Clearly
${J_{t}}^{\mu}{J_{\phi}}_{\mu}=0$
and when $J_{\phi}$
is spacelike $J_{t}$ is timelike and viceversa.

\begin{figure}\centering
\includegraphics[scale=0.3]{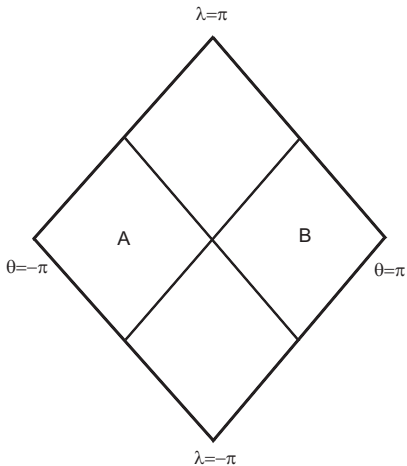}
\caption{\label{F:btzpoincare}\footnotesize{This figure shows how
the covering space of BTZ on the boundary is related to the
Poincar\'{e} chart. The big diamond is the Poincar\'{e} chart
whereas the covering space of BTZ is the A and B small diamonds.}}
\end{figure}

Now let us see what the region of the boundary of CAdS covered by
the Poincar\'{e} chart is. Here we are going to consider just one
fundamental region of CAdS, $\lambda\in[-\pi,\pi)$. From
(\ref{E:105}) and (\ref{E:107}) it follows that
\begin{equation}\label{E:25}
\cos\lambda=\frac{z^{2}+\ell^{2}+k^{2}-T^{2}}{\sqrt{\left(z^{2}+\ell^{2}+k^{2}-T^{2}\right)^{2}+\left(2\ell
T\right)^{2}}}
\end{equation}
and
\begin{equation}\label{E:26}
\cos\theta=-\frac{z^{2}-\ell^{2}+k^{2}-T^{2}}{\sqrt{\left(z^{2}+\ell^{2}+k^{2}-T^{2}\right)^{2}+\left(2\ell
T\right)^{2}-\left(2\ell z\right)^{2}}}.
\end{equation}
By following the analysis in \cite{caBallrfBra07} it can be shown
that the equality $\cos\lambda=-\cos\theta$ can be satisfied on
the surface $\rho=\pi/2$ and it corresponds to, let us say, the
boundary of one Poincar\'{e} chart in the boundary, see figure
\ref{F:btzpoincare}. We see that the covering region of the
maximally extended BTZbh is half of the Poincar\'{e} chart. Let us
see what the relation between BTZ coordinates and Poincar\'{e}
coordinates is.

From (\ref{E:107}) and (\ref{E:112}) it follows that in the boundary
\begin{equation}\label{E:116}
T=\ell e^{-\ell\kappa\phi}\sinh\kappa t\hspace{1cm}k=\ell
e^{-\ell\kappa\phi}\cosh\kappa t.
\end{equation}
From (\ref{E:116}) we see that the relation between BTZ
coordinates and Poincar\'{e} coordinates is analogous to the
relation between Rindler and Minkowski coordinates
\cite{ndBpcwD82}. This fact suggests that some kind of Unruh
effect is taking place in the boundary of the CAdS spacetime. In
the next sections we shall show this is indeed the case.


The vector fields of $\partial_{t}$ and $\partial_{\phi}$ for the
non-rotating 
case are given in figure 4\footnote{Similar plots have been given before in
\cite{sAminiBensHol99} and \cite{sAminiBen08}.}.

\begin{figure}\centering
\includegraphics[scale=0.4]{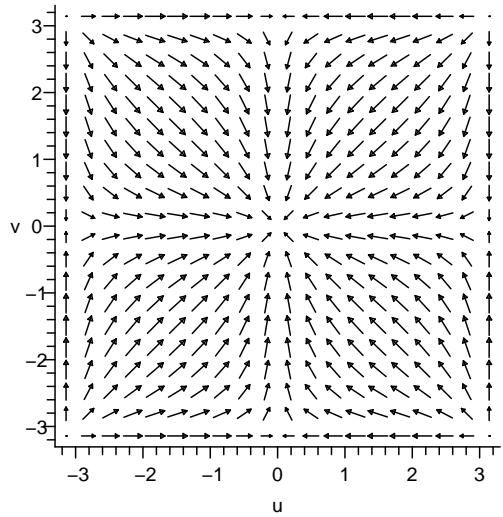}
\includegraphics[scale=0.4]{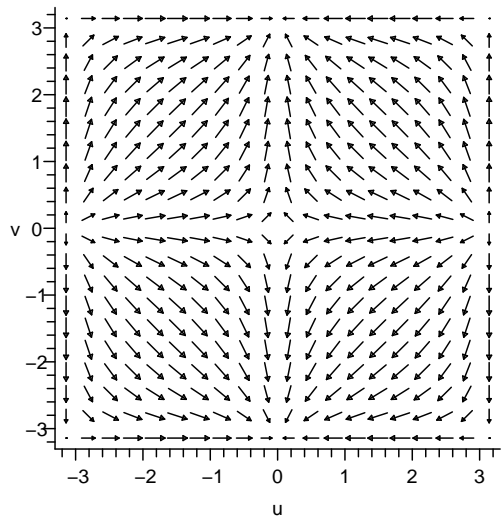}
\caption{\label{F:fieldnorotating}\footnotesize{These figures show
the vector fields $\partial_{\phi}$ and $\partial_{t}$
respectively for the nonrotating BTZ black hole. These figures
were made with Maple 10.}}
\end{figure}


\section{Wedge regions in AdS spacetime}

In the context of Algebraic Holography \cite{khReh00} wedge
regions in AdS spacetime play a prominent r\^{o}le. In this
section we will show that the exterior of the BTZbh is a wedge
region.

Following \cite{khReh00} we define a wedge region in AdS spacetime
as follows. Let us take two light-like vectors ($e,f$) in the
embedding space $\mathbb{R}^{2,2}$ such that $e\cdot f>0$, then a
wedge region in AdS spacetime is defined by
\begin{equation}\label{E:117}
\widetilde{W}(e,f)=\{X\in\mathbb{R}^{2+2}:X^{2}=-\ell^{2}, e\cdot
X<0,f\cdot X<0\}.
\end{equation}
This region has two connected components. One where the resulting
vector by acting on the tangent vector at the point
$X\in\widetilde{W}(e,f)$ with the boost in the $e$-$f$ plane is
future directed and other where it is past directed. This is a
consequence of the fact that the vector $\delta_{e,f}X=(f\cdot
X)e-(e\cdot X)f$ is time-like, $(\delta_{e,f}X)^{2}=-2(e\cdot
f)(e\cdot X)(f\cdot X)<0$. The wedges regions are defined in
\cite{khReh00} as these regions module the identification
$X\leftrightarrow-X$\footnote{It is worth to note that a more
primitive definition of a wedge region in AdS spacetime is to
define it as the intersection of AdS spacetime with a wedge region
in the embedding space.}. Hence in order to check that a region in
AdS spacetime is a wedge region we just have to verify that it has
these properties. Let us do this for the exterior of the BTZbh.

\subsection{The exterior of the BTZ black hole is a wedge region}

Let us take $e=(1,0,0,-1)$ and $f=(-1,0,0,-1)$. These light-like
vectors satisfy $e\cdot f>0$. By using the parametrization
(\ref{E:112})-(\ref{E:113a}) we get
\begin{equation}\label{E:118}
e\cdot X=-\sqrt{B(r)}\left(\sinh\tilde{t}+\cosh\tilde{t}\right)<0
\end{equation}
and
\begin{equation}\label{E:119}
f\cdot X=\sqrt{B(r)}\left(\sinh\tilde{t}-\cosh\tilde{t}\right)<0.
\end{equation}
Hence the exterior of the rotating BTZ black hole is a wedge
region. It is also true for the non-rotating case. As explained in
\cite{khReh00} this wedge intersects the boundary in a double
cone. In the previous section we have found this double cone
explicitly. These regions are preserved by the action of the
subgroup generated by the Killing vector $\partial_{t}$. This
group acts on the wedge as a subgroup of the AdS group and as
subgroup of the conformal group on the boundary.

\subsection{The exterior of BTZ black hole and the Rindler wedge in Minkowski spacetime}

We have found that the exterior of BTZ black hole is invariant
under the one-parameter subgroup of the AdS group generated by
$\partial_{t}$. Let us calculate explicitly this one-parameter
subgroup. The generator of this subgroup is given by (\ref{E:20}),
hence a matrix representation of it on the vector space
$\mathbb{R}^{2,2}$ is given by
\begin{equation}\label{E:120}
\partial_{t}=\kappa\left( \begin{array}{cccc}
0 & 0 & 0 & 1 \\
0 & 0 & 0 & 0 \\
0 & 0 & 0 & 0 \\
1 & 0 & 0 & 0 \end{array} \right).
\end{equation}
The matrix (\ref{E:120}) is an element of the Lie
algebra of $SO_{0}(2,2)$ and the one-parameter subgroup generated
by it is given by
\begin{equation}\label{E:123}
\Lambda(t)=e^{t\partial_{t}}=\left( \begin{array}{cccc}
\cosh\kappa t & 0 & 0 & \sinh\kappa t \\
0 & 1 & 0 & 0 \\
0 & 0 & 1 & 0 \\
\sinh\kappa t & 0 & 0 & \cosh\kappa t \end{array} \right).
\end{equation}
This one-parameter group acts on the vector space
$\mathbb{R}^{2,2}$ leaving the exterior of the BTZbh invariant.

As we said in section 2, AdS spacetime has a compactified
Minkowski spacetime at infinity. Also we know that $SO_{0}(2,2)$
acts as the conformal group on this Minkowski spacetime
\cite{rHaag96}. Let us see to which element of the conformal group
in two dimensions the element (\ref{E:123}) corresponds.


Remembering the definition of the coordinates of Minkowski
spacetime at infinity we have
\begin{equation}\label{E:125}
\xi^{1}=\frac{u}{v+x}\hspace{1cm}\xi^{2}=\frac{y}{v+x}.
\end{equation}
If now we let $\Lambda(t)$ to act on $X^{T}=(u,v,x,y)$ we obtain
\begin{equation}\label{E:126}
\Lambda(t)x=\left( \begin{array}{c}
u\cosh\kappa t+y\sinh\kappa t \\
v \\
x \\
u\sinh\kappa t +y\cosh\kappa t \end{array} \right)=\left(
\begin{array}{c}
u' \\
v' \\
x' \\
y' \end{array}\right).
\end{equation}
This transformation on the null cone, $\mathcal{C}^{4}$, induces a
transformation on $\xi^{1}$ and $\xi^{2}$.
\begin{equation}\label{E:127}
\xi^{1}\rightarrow\xi'^{1}=\frac{u'}{v'+x'}\hspace{1cm}\xi^{2}\rightarrow\xi'^{2}=\frac{y'}{v'+x'}.
\end{equation}
From (\ref{E:126}) and (\ref{E:125}) it follows that
\begin{equation}\label{E:128}
\xi'^{1}=\xi^{1}\cosh\kappa t+\xi^{2}\sinh\kappa
t\hspace{1cm}\xi'^{2}=\xi^{1}\sinh\kappa t+\xi^{2}\cosh\kappa t.
\end{equation}
Then the subgroup of the AdS group generated by $\partial_{t}$
corresponds to a Lorentz boost in the two dimensional Minkowski
spacetime. From this we can see that the Rindler wedge in this two
dimensional Minkowski spacetime is invariant under the action of
the subgroup of the global conformal group corresponding to the
subgroup of the AdS group generated by $\partial_{t}$. The Rindler wedge just mentioned is the double cone which results from the intersection
of the wedge corresponding to the exterior of the BTZ black hole with the boundary.

The correspondence between the others one-parameter subgroups can
be found analogously. For example let us analyze the subgroup
generated by $\partial_{\phi}$.

The matrix representation of this generator on the vector space
$\mathbb{R}^{2,2}$ is given by
\begin{equation}\label{E:129}
\partial_{\phi}=\kappa \ell\left( \begin{array}{cccc}
0 & 0 & 0 & 0 \\
0 & 0 & 1 & 0 \\
0 & 1 & 0 & 0 \\
0 & 0 & 0 & 0 \end{array} \right).
\end{equation}
Hence the finite transformation is given by
\begin{equation}\label{E:130}
\Lambda(\phi)=e^{\phi\partial_{\phi}}=\left( \begin{array}{cccc}
1 & 0 & 0 & 0 \\
0 & \cosh\kappa \ell\phi & \sinh\kappa \ell\phi & 0 \\
0 & \sinh\kappa \ell\phi & \cosh\kappa \ell\phi & 0 \\
0 & 0 & 0 & 1 \end{array} \right).
\end{equation}
If we let this transformation to act on $X^{T}=(u,v,x,y)$ we
obtain
\begin{equation}\label{E:131}
\Lambda(\phi)x=\left( \begin{array}{c}
u \\
v\cosh\kappa \ell\phi+x\sinh\kappa \ell\phi \\
v\sinh\kappa \ell\phi+x\cosh\kappa \ell\phi \\
y \end{array} \right)=\left(
\begin{array}{c}
u' \\
v' \\
x' \\
y' \end{array}\right).
\end{equation}
This transformation induces a transformation on $\xi^{1}$ and
$\xi^{2}$ given by
\begin{equation}\label{E:132}
\xi'^{1}=e^{-\kappa
\ell\phi}\xi^{1}\hspace{1cm}\xi'^{2}=e^{-\kappa \ell\phi}\xi^{1}.
\end{equation}
Then the subgroup of AdS group generated by $\partial_{\phi}$
corresponds to the dilation group on the two dimensional Minkowski
spacetime.

\section{Thermal state in AdS spacetime and in the BTZ black hole}

In this section we show that there exists an equilibrium thermal
state in AdS spacetime in 1+2 dimensions and discuss its relation
to an equilibrium thermal state in the BTZbh.

In the previous section we showed that the exterior of the BTZbh
is a wedge region. Now, also the Poincar\'{e} chart is likely to
be a wedge region. If so we can associate a net of algebras to
these regions. This can be done, for example, by adapting the
formalism introduced in \cite{jDim80} to the present case. Due to
the invariance of these wedges regions under the action of the
subgroups generated by $\partial_{T}$ and $\partial_{t}$\footnote{Here $\partial_{T}$ and $\partial_{t}$ are the Killing vectors associated with translations in $T$ and $t$ respectively.} we have
\begin{equation}\label{E:132b}
\omega(\alpha_{T}a)=\omega(a)\hspace{1cm}a\in A(W_{T})
\end{equation}
and
\begin{equation}\label{E:132c}
\omega(\alpha_{t}a)=\omega(a)\hspace{1cm}a\in A(W_{t}),
\end{equation}
where $W_{T}$ and $W_{t}$ are the wedge regions associated to the
Poincar\'{e} chart and the exterior of the BTZbh respectively, and $A$ is a von Neumann algebra\footnote{Here we are following the conditions on the algebra $A$ used in Algebraic Holography \cite{khReh00}. The conclusions we get are as valid as Algebraic Holography is.}. The
symbol $\omega$ denote a state\footnote{This state should satisfy certain mathematical conditions, see for example \cite{rHaag96} p. 122.} on these algebras, below we explain
more about this state. The symbols $\alpha_{T}$ and
$\alpha_{t}$ denote the automorphisms of the algebras associated
to $W_{T}$ and $W_{t}$ respectively. We assume that these
automorphisms satisfy
\begin{equation}\label{E:132a}
\alpha_{T}A(W_{T})=A(\Lambda(T)W_{T})
\end{equation}
and
\begin{equation}\label{E:132aa}
\alpha_{t}A(W_{t})=A(\Lambda(t)W_{t})
\end{equation}
where $\Lambda(T)$ and $\Lambda(t)$ are the transformations which
leave invariant $W_{T}$ and $W_{t}$ respectively. The last four
expressions deserve some comments. The existence of $\Lambda(T)$
and $\Lambda(t)$ is a consequence of the existence of the Killing
vectors $\partial_{T}$ and $\partial_{t}$, which is a geometrical
property of AdS spacetime. The equations (\ref{E:132a}) and
(\ref{E:132aa}) are part of the assumptions about the structure of
the algebras. The equations (\ref{E:132b}) and (\ref{E:132c}) are
a consequence of analogous expressions in the boundary assuming
AH. Hence once we are in AdS spacetime and its geometry and we
postulate the algebraic structure on its boundary these four
equations should be valid. Now let us make some comments about the
state $\omega$. As we have said the last four equations have a
bulk and boundary counterpart. In the boundary we have a Minkowski
spacetime whereas in the bulk a spacetime with constant curvature.
If we want to make quantum field theory on both and both should be
equivalent there should be no preference for one of these two
perspectives \textit{a priori}. However if we take into account
that Quantum Field Theory in Minkowski spacetime has a well
establish theory it seems that we should go from the boundary to
the bulk, because in this way we can use all the formalism at hand
for QFT in Minkowski spacetime and try to apply it to the bulk of
AdS spacetime. It could be possible that some tools for Minkowski
spacetime do not apply to AdS spacetime however we will notice
this if we get a contradiction or an unphysical result. By taking
this philosophy it seems that we should take the ground state on
the boundary defined with respect to $\partial_{T}$ as the vacuum
of the theory on the boundary. If we do this then in the bulk
$\omega$ will be our vacuum, i.e., if we make the GNS construction
\cite{rHaag96} of this state then the unitary operator associated
with translations in $T$ has a self-adjoint generator operator,
$H_{T}$, with spectrum $[0,\infty]$. This hamiltonian satisfies
\begin{equation}\label{E:132aaa}
H_{T}|\Psi_{\omega}\rangle=0,
\end{equation}
where $|\Psi_{\omega}\rangle$ is the cyclic vector associated with
$\omega$ through the GNS construction. Also we have assumed the
unitary operator implementing translation in $T$ is strongly
continuos with respect to $T$. Hence in this case the Von
Neumann's theorem \cite{mReedbSim80} assures the existence of
$H_{T}$.

As was proven in \cite{khReh00}, once we have set up this scenario
we can apply the well-known theorems of Takesaki-Tomita and
Bisognano-Wichmann to the theory on the boundary. From this it follows that the vacuum in the boundary becomes an equilibrium thermal state with respect to $t$ when restricted to the double cone associated with the exterior region of the BTZbh. This is on the boundary. Now, by applying Algebraic Holography we obtain that the same is valid in the bulk, so $\omega$ in the bulk is also an equilibrium equilibrium thermal state with respect to $t$ when restricted to the exterior of the BTZbh. Put in
other way, it satisfies the KMS condition with respect to $t$\footnote{For a similar result in the Schwarzschild black hole see \cite{gSew80}.}. Let
us find what the temperature of this state is.

From (\ref{E:126}) we can see that the parameter of the boost is
$t'=\kappa t$ with $\kappa=\frac{r_{+}}{\ell^{2}}$. Using the
theorem 4.1.1 (Bisognano-Wichmann theorem) in \cite{rHaag96} we
have that the parameter of the modular group which appears in the
Tomita-Takesaki theorem is given by
\begin{equation}\label{E:133}
\tau=-\frac{t'}{2\pi}=-\frac{\kappa}{2\pi}t.
\end{equation}
Using the theorem 2.1.1 (Tomita-Takesaki theorem) in
\cite{rHaag96} it is possible to proof the state invariant under
the modular group satisfies the KMS condition with
$\beta=-1=\frac{1}{\textsf{T}}$, see \cite{rHaag96} p. 218
\begin{equation}\label{E:134}
\omega\left(\left(\alpha_{\tau}A\right)B\right)=\omega\left(B\left(\alpha_{\tau-i}A\right)\right).
\end{equation}
Hence from (\ref{E:133}) it follows the temperature of the thermal
state with respect to $t$ is
\begin{equation}\label{E:135}
\textsf{T}=\frac{\kappa}{2\pi}.
\end{equation}
This is the so-called temperature of the black hole. The local
temperature measure by an observer at constant radius $r$ is
\begin{equation}\label{E:135a}
\textsf{T}(r)=\frac{1}{(-g_{00})^{1/2}}\frac{\kappa}{2\pi}.
\end{equation}
This is because the proper time of this observer, $\lambda$, and
the time $t$ are related as $\lambda=(-g_{00})^{1/2}t$.

So far we have shown that $\omega$ satisfies the KMS condition on
the covering space of one exterior of the BTZbh, however the
exterior of the BTZbh is obtained after making $\phi$
2$\pi$-periodic. This periodicity introduces new features because
we have a non simply connected spacetime, a cylinder, instead of a
simply connected spacetime, a plane. We shall assume that there is a way to construct the
thermal state on the cylinder from one on the plane algebraically.
Let us call the state on the covering space of one exterior region
of the BTZbh $\omega_{AdS}$ and on the exterior region of the
BTZbh $\omega_{BTZ}$.

The state $\omega_{BTZ}$ is defined in
$\mathbb{R}^{1}\times\mathbb{S}^{1}$ which is one exterior region
of the BTZ black hole whereas $\omega_{AdS}$ is defined on
$\mathbb{R}^{1}\times\mathbb{R}^{1}$ which is the covering space
of one exterior region of the BTZ black hole. Put in this way, the
state $\omega_{BTZ}$ is a thermal state on a black hole, i.e., the
Hawking effect for an eternal black hole takes place and
corresponds to the Unruh effect on AdS after making $\phi$
2$\pi$-periodic. Put in this form we can say that the Hawking
effect in the eternal BTZbh has its origin in the Unruh effect in
the boundary of AdS spacetime and in the topological relation
between $\mathbb{R}^{1}\times\mathbb{R}^{1}$ and
$\mathbb{R}^{1}\times\mathbb{S}^{1}$.

In the rotating case there is a tiny modification in the analysis.
From the form of the parametrization of the exterior of BTZ black
hole (\ref{E:112}) we see that in the rotating case the subgroup
of the AdS group which leaves invariant the wedge $W$ is generated
by $\partial_{\tilde{t}}$. Following the analysis of the previous
section, now the parameter of the modular group which appears in
Tomita-Takesaki theorem is related to the time $\tilde{t}$ as
\begin{equation}\label{E:138}
\tau=-\frac{\tilde{t}}{2\pi}.
\end{equation}
Hence the state is thermal with respect to $\tilde{t}$ at
temperature $\textsf{T}=\frac{1}{2\pi}$. Put in this form, because
the state $\omega$ satisfies
\begin{equation}\label{E:139}
\omega\left(\left(\alpha_{\tilde{t}-2\pi}A\right)B\right)=\omega\left(B\left(\alpha_{\tilde{t}}A\right)\right),
\end{equation}
then there is a periodicity of the state in $t$ and $\phi$ given
by
\begin{equation}\label{E:140}
(t, \phi)\rightarrow (t-i\beta,
\phi+i\Omega_{H}\beta)\hspace{1cm}\beta=\frac{1}{\textsf{T}}=\frac{2\pi}{\kappa}.
\end{equation}
where $\beta=\frac{2\pi}{\kappa}$. Hence again the black hole is
hot at temperature $\textsf{T}=\frac{\kappa}{2\pi}$.

\subsection{Thermal state for a real quantum scalar field}

In this section we show how an equilibrium thermal state in the
bulk of AdS spacetime maps to an equilibrium thermal state on its
boundary for a real quantum scalar field.

The equation the field satisfies is
\begin{equation}\label{E:140a}
\left(\nabla_{\mu}\nabla^{\mu}-\xi R-m^{2}\right)\phi=0,
\end{equation}
where $\xi$ is a coupling constant, $R$ is the Ricci scalar and
$m$ can be considered as the mass of the field. For the metric of
AdS spacetime, $R=6\Lambda=-\frac{6}{\ell^{2}}$. Hence the last
equation can be written as
\begin{equation}\label{E:140bb}
\left(\nabla_{\mu}\nabla^{\mu}-\widetilde{m}^{2}\right)\phi=0,
\end{equation}
where $\widetilde{m}^{2}=m^{2}-\frac{6\xi}{\ell^{2}}$.
For our purposes it is convenient to use Poincar\'{e} coordinates.
In these coordinates $g=-\frac{\ell^{6}}{z^{6}}$. Because
$\partial_{T}$ and $\partial_{k}$ are Killing vectors we propose
the ansatz $\phi(T,k,z)\propto e^{-i\omega T}e^{ink}f_{\omega
n}(z)$.

The normalized modes turns out to be \cite{lOrtphd11}
\[
F_{na}(T,k,z)=\left(\frac{a}{4\ell\pi\omega}\right)^\frac{1}{2}e^{-i\omega
T}e^{ink}zJ_{p}(az).
\]

It follows that the two point function is \cite{lOrtphd11}
\begin{eqnarray}\label{E:140kk}
F(\Delta T,\Delta k;z,z')&=&
\frac{(zz')^{1+p}}{2\ell\pi}\lim_{\epsilon\rightarrow
0_{+}}\Bigg\{\\
&&\frac{F_{4}\left(\alpha,\alpha;\alpha,\alpha;-\frac{z^{2}}{(\Delta
k)^{2}-(\Delta T+i\epsilon)^{2}},-\frac{z'^{2}}{(\Delta
k)^{2}-(\Delta T+i\epsilon)^{2}}\right)}{((\Delta k)^{2}-(\Delta
T+i\epsilon)^{2})^{1+p}}\Bigg\}\nonumber
\end{eqnarray}
where $\alpha=1+p$, $p^{2}=1+\ell^{2}\widetilde{m}^{2}$ and
$F_{4}$ is a Hypergeometric function of two variables. Following
\cite{mBerjBrosuMosrSch00} we now make $z=z'$ and multiply
(\ref{E:140kk}) by $z^{-2(1+p)}$ and take the limit $z\rightarrow
0$ we obtain
\begin{eqnarray}\label{E:140ll}
F_{b}(\Delta T,\Delta k)&\equiv &\lim_{z\rightarrow
0}z^{-2(1+p)}F(\Delta T,\Delta k;z,z)\nonumber\\
&=&\frac{1}{2\ell\pi}\lim_{\epsilon\rightarrow
0_{+}}\frac{1}{((\Delta k)^{2}-(\Delta T+i\epsilon)^{2})^{1+p}}.
\end{eqnarray}

Now let us analyze what happens when we restrict (\ref{E:140ll})
to the exterior of BTZ black hole. By using (\ref{E:116}) we get
\begin{equation}\label{E:140oo}
F_{b}(\Delta t,\Delta
\phi')=\frac{1}{2\ell\pi}\frac{(2\ell^{2}e^{-\kappa(\phi_{1}'+\phi_{2}')})^{-1-p}}{(\cosh\kappa\Delta\phi'-\cosh\kappa\Delta
t-i\epsilon)^{1+p}},
\end{equation}
where $\phi'=\ell\phi$.

From (\ref{E:140oo}) it follows that
\begin{equation}\label{E:140ooooa}
F_{b}(-\Delta t,\Delta \phi')=F_{b}(\Delta t-i\beta,\Delta \phi')
\end{equation}
where $\beta=\frac{1}{\textsf{T}}=\frac{2\pi}{\kappa}$. From
(\ref{E:140ooooa}) it follows that the restricted two point
function to the exterior of the BTZbh in the boundary satisfies
the KMS condition \cite{saFullsnRui87}. Hence (\ref{E:140kk}) is a
thermal state at temperature $\textsf{T}=\frac{\kappa}{2\pi}$ when
restricted to the exterior of the BTZbh. From (\ref{E:140kk}) we
see that the thermal property does not change when we take the
limit $z\rightarrow 0$, since $z=\ell
e^{-\kappa\phi'}\frac{r_{+}}{r}$, hence we can say the thermal
state in the bulk of AdS maps to a thermal state on its boundary,
and this state is given by (\ref{E:140oo}).

The Jacobian of (\ref{E:116}) is
\begin{equation}\label{E:140pp}
\left|\frac{\partial(T,k)}{\partial(t,\phi')}\right|=\ell^{2}\kappa^{2}e^{-2\kappa\phi'}.
\end{equation}
If we consider (\ref{E:116}) as a conformal transformation then
from (\ref{E:140pp}) and the relationship between two conformal
metrics
\begin{equation}\label{E:140qq}
g_{\mu\nu}'(x')=\Omega^{2}(x)g_{\mu\nu}(x)
\end{equation}
it follows that in the present case
\begin{equation}\label{E:140rr}
\Omega(\phi')=\frac{1}{\ell\kappa e^{-\kappa\phi'}}.
\end{equation}
If we want to consider $F_{b}$ as a correlation function for a
conformal field theory in the boundary of AdS spacetime and take
into account two correlation function in a conformal field theory
are related by \cite{pdFranpMatdSene97} p. 104
\begin{equation}\label{E:140ss}
\Omega(x_{1}')^{-\Delta_{1}}\Omega(x_{2}')^{-\Delta_{2}}\langle\phi_{1}(x_{1}')\phi_{2}(x_{2}'))\rangle=\langle\phi_{1}(x_{1})\phi_{2}(x_{2})\rangle,
\end{equation}
where $\Delta_{1}$ and $\Delta_{2}$ are the scaling dimension of
the field $\phi_{1}$ and $\phi_{2}$ respectively, then from the
equality
\begin{equation}\label{E:140tt}\nonumber
\frac{1}{2\pi}\frac{\Omega(\phi_{1}')^{-1-p}\Omega(\phi_{2}')^{-1-p}}{((\Delta
k)^{2}-(\Delta
T+i\epsilon)^{2})^{1+p}}=\frac{2^{-1-p}}{2\pi}\frac{\kappa^{2(1+p)}}{(\cosh\kappa\Delta\phi'-\cosh\kappa\Delta
t-i\epsilon)^{1+p}}
\end{equation}
we finally have
\begin{equation}\label{E:140uu}
F_{b}(\Delta t,\Delta \phi')\simeq\frac{1}{2\ell\pi
2^{(1+p)}}\frac{\kappa^{2(1+p)}}{(\cosh\kappa\Delta\phi'-\cosh\kappa\Delta
t-i\epsilon)^{(1+p)}}.
\end{equation}
According to standard conformal field theory
\cite{pdFranpMatdSene97} this correlation function would
correspond to a field with scaling dimension $\Delta=1+p$.

From (\ref{E:140uu}) we can obtain the two correlation functions
on the exterior of BTZ black hole and its covering space
respectively by using the image method. Let us explain briefly
this method. The image sum method relates the two point function
in a multiple connected and the two point function in a simply
connected spacetime, see \cite{rBasDow79} or \cite{crCrambsKay96}.
In this case we could apply the formula
\begin{equation}\label{E:142b}
F'_{b}(\Delta t,\Delta\phi)=\sum_{n\in\mathbb{Z}}e^{2\pi
in\alpha}F_{b}(\Delta t,\Delta\phi+2\pi n),
\end{equation}
where $F'_{b}(\Delta t,\Delta\phi)$ is the two point function in
the covering space of the exterior of BTZ black hole and $\alpha$
a parameter which is $0$ for untwisted and $\frac{1}{2}$ for
twisted fields. In this work we are interested in untwisted
fields, so we take $\alpha=0$. Applying this method to
(\ref{E:140uu}) we obtain
\begin{eqnarray}\label{E:140vv}
F'_{b}(\Delta
t,\Delta\phi)&\equiv&\sum_{n\in\mathbb{Z}}\frac{1}{2\ell\pi
2^{(1+p)}}\times\nonumber\\
&\times&\frac{\kappa^{2(1+p)}}{(\cosh\kappa\ell(\Delta\phi+2\pi
n)-\cosh\kappa\Delta t-i\epsilon)^{(1+p)}}.
\end{eqnarray}
This correlation function would correspond to fields defined on
the same covering space of one exterior region of BTZ black hole.
However BTZ black hole has two exterior regions analogously to
Schwarzschild black hole. The parametrization of the other
exterior region is given by changing the sign in (\ref{E:116}).
Hence in this case the correlation function is
\begin{eqnarray}\label{E:140vvv}
F'_{b}(\Delta
t,\Delta\phi)&\equiv&\sum_{n\in\mathbb{Z}}\frac{1}{2\ell\pi
2^{(1+p)}}\times\nonumber\\
&\times &\frac{\kappa^{2(1+p)}}{(\cosh\kappa\ell(\Delta\phi+2\pi
n)+\cosh\kappa\Delta t+i\epsilon)^{(1+p)}}.
\end{eqnarray}

From the previous analysis is clear that on the boundary we have a
thermal state when we restrict the state defined on the
Poincar\'{e} chart to the covering space of one exterior region of
the BTZbh. Also because we have to make $\phi$ $2\pi$-periodic
then this covering space is a cylinder with spatial cross section.
Now a massless field theory on a cylinder is equivalent to a two
dimensional conformal field theory \cite{pdFranpMatdSene97}, hence
we can say that on the boundary of AdS we have two conformal
theories related in such way that when we restrict a vacuum state
to one of them it becomes a thermal state. We point out that the
expressions (\ref{E:140vv}) and (\ref{E:140vvv}) have been given
in \cite{jMalda01} and according to it they have been obtained in
\cite{eKe99}. In \cite{eKe99} it was used the proposal given in
\cite{eWitt98} for the AdS/CFT correspondence. However we have
just obtained them by using AdS/CFT in QFT which uses both
Algebraic Holography and the boundary-limit Holography. This last
procedure can be considered as alternative to the procedure used
by Keski-Vakkuri\footnote{See also \cite{bKaylOrt11}.}.


So far in this section we have considered the non rotating BTZ
black hole. However similar considerations apply to the rotating
case. In the rotating case the relation between Poincar\'{e} and
BTZ coordinates is
\begin{equation}\label{E:140xx}
T=\pm
\ell\left(\frac{r^{2}-r_{+}^{2}}{r^{2}-r_{-}}\right)^{1/2}e^{-\tilde{\phi}}\sinh\tilde{t}\hspace{1cm}k=\pm
\ell\left(\frac{r^{2}-r_{+}^{2}}{r^{2}-r_{-}}\right)^{1/2}e^{-\tilde{\phi}}\cosh\tilde{t},
\end{equation}
where $\tilde{t}$ and $\tilde{\phi}$ are defined in
(\ref{E:113a}). Hence in the limit $r\rightarrow\infty$
\begin{equation}\label{E:140yy}
T=\pm \ell e^{-\tilde{\phi}}\sinh\tilde{t}\hspace{1cm}k=\pm \ell
e^{-\tilde{\phi}}\cosh\tilde{t}.
\end{equation}
The sign $+$ corresponds to one exterior of BTZ black hole and the
sign $-$ to the other. Following the analysis for the non-rotating
case we have
\begin{equation}\label{E:140oooo}
F_{b}(-\Delta\tilde{t},\Delta \tilde{\phi})=F_{b}(\Delta
\tilde{t}-i2\pi,\Delta \tilde{\phi})
\end{equation}
if
\begin{equation}\label{E:140bbb}
(t, \phi)\rightarrow (t-i\beta, \phi+i\beta\Omega_{H})
\end{equation}
where $\beta=\frac{1}{\textsl{T}}$ and $\Omega_{H}$ is the angular
velocity of the horizon.

\section{Summary and open questions}

In this work, apart from reviewing the relationship between AdS
spacetime and the BTZ black hole, we studied how a ground state
(the Poincar\'{e} vacuum) in AdS spacetime in 1+2 dimensions
becomes a thermal state in the BTZ black hole. We address this
issue in the abstract setting of Algebraic Holography and in a
concrete example of the quantum real scalar field. We found both
approaches consistent and we can say that the Unruh effect in AdS
spacetime becomes the Hawking effect for the eternal BTZ black
hole. In doing this we obtained that the thermal state in the BTZ
black hole maps to its boundary. This state coincides with the
state obtained in \cite{eKe99} by a rather different methods based
on the proposal \cite{eWitt98} for the AdS/CFT correspondence. So,
a natural question we can ask ourselves is to find out the reason
of this coincidence. Is it accidental? Or, what is the reason in
this case the two approaches \cite{mBerjBrosuMosrSch00} and
\cite{eWitt98} give the same result? We leave these questions open
for now and hope to solve them in future work.

\vspace{0.5cm}

Acknowledgments: I thank my supervisor, Dr. Bernard S. Kay, for
suggesting me to study equilibrium thermal states in AdS spacetime
and their relation to equilibrium thermal states in the BTZbh. I
also thank him for his guidance and helpful advise during this
work.

This work was carried out with the sponsorship of CONACYT Mexico
grant 302006.

\appendix

\section{Global conformal transformations in two dimensions}

In section 5 we found explicitly the subgroups of the global
conformal group in two dimensions generated by $J_{uy}$ and
$J_{vx}$. In this appendix we give the others subgroups of this
group.

First let us remember the elements of the Lie algebra of the AdS
group. These elements are
\begin{equation}\label{E:142a}
\begin{array}{cc}
J_{uv}=u\partial_{v}-v\partial_{u} &
\quad\quad J_{xy}=x\partial_{y}-y\partial_{x} \\
J_{ux}=u\partial_{x}+x\partial_{u} & \quad\quad
J_{uy}=u\partial_{y}+y\partial_{u} \\
J_{vx}=v\partial_{x}+x\partial_{v} & \quad\quad
J_{vy}=v\partial_{y}+y\partial_{v}
\end{array}
\end{equation}

It is well known that in a representation on the vector space
$\mathbb{R}^{2,2}$ these generators can be represented by the
matrices
\begin{equation}\label{E:142b}
J_{uv}=\left( \begin{array}{cccc}
0 & -1 & 0 & 0 \\
1 & 0 & 0 & 0 \\
0 & 0 & 0 & 0 \\
0 & 0 & 0 & 0 \end{array} \right)\hspace{1cm}J_{xy}=\left(
\begin{array}{cccc}
0 & 0 & 0 & 0 \\
0 & 0 & 0 & 0 \\
0 & 0 & 0 & -1 \\
0 & 0 & 1 & 0 \end{array} \right)
\end{equation}
\begin{equation}\label{E:142c}
J_{ux}=\left( \begin{array}{cccc}
0 & 0 & 1 & 0 \\
0 & 0 & 0 & 0 \\
1 & 0 & 0 & 0 \\
0 & 0 & 0 & 0 \end{array} \right)\hspace{1cm}J_{uy}=\left(
\begin{array}{cccc}
0 & 0 & 0 & 1 \\
0 & 0 & 0 & 0 \\
0 & 0 & 0 & 0 \\
1 & 0 & 0 & 0 \end{array} \right)
\end{equation}
\begin{equation}\label{E:142d}
J_{vx}=\left( \begin{array}{cccc}
0 & 0 & 0 & 0 \\
0 & 0 & 1 & 0 \\
0 & 1 & 0 & 0 \\
0 & 0 & 0 & 0 \end{array} \right)\hspace{1cm}J_{vy}=\left(
\begin{array}{cccc}
0 & 0 & 0 & 0 \\
0 & 0 & 0 & 1 \\
0 & 0 & 0 & 0 \\
0 & 1 & 0 & 0 \end{array} \right)
\end{equation}

For our purposes the relevant elements of the Lie algebra of the
AdS group are
\begin{equation}\label{E:143}
A=J_{ux}-J_{uv}=\left( \begin{array}{cccc}
0 & 1 & 1 & 0 \\
-1 & 0 & 0 & 0 \\
1 & 0 & 0 & 0 \\
0 & 0 & 0 & 0 \end{array} \right),
\end{equation}
\begin{equation}\label{E:144}
B=J_{xy}+J_{vy}=\left( \begin{array}{cccc}
0 & 0 & 0 & 0 \\
0 & 0 & 0 & 1 \\
0 & 0 & 0 & -1 \\
0 & 1 & 1 & 0 \end{array} \right),
\end{equation}
\begin{equation}\label{E:145}
C=J_{uv}+J_{ux}=\left( \begin{array}{cccc}
0 & -1 & 1 & 0 \\
1 & 0 & 0 & 0 \\
1 & 0 & 0 & 0 \\
0 & 0 & 0 & 0 \end{array} \right),
\end{equation}
\begin{equation}\label{E:146}
D=J_{xy}-J_{vy}=\left( \begin{array}{cccc}
0 & 0 & 0 & 0 \\
0 & 0 & 0 & -1 \\
0 & 0 & 0 & -1 \\
0 & -1 & 1 & 0 \end{array} \right)
\end{equation}

Let $\textbf{a}=(a,b)$ be a two dimensional vector. Then
\begin{equation}\label{E:147}
\Lambda(\textbf{a})=e^{aA+bB}=\left( \begin{array}{cccc}
1 & a & a & 0 \\
-a & 1+\frac{b^{2}-a^{2}}{2} & \frac{b^{2}-a^{2}}{2} & b \\
a & \frac{a^{2}-b^{2}}{2} & 1+\frac{a^{2}-b^{2}}{2} & -b \\
0 & b & b & 1
\end{array} \right).
\end{equation}
If we apply this transformation to $X^{T}=(u,v,x,y)$ we get
\begin{equation}\label{E:148}
\left( \begin{array}{c}
u' \\
v' \\
x' \\
y'
\end{array} \right)=\left( \begin{array}{cccc}
u+av+ax \\
-au+\left(1+\frac{b^{2}-a^{2}}{2}\right)v + \left(\frac{b^{2}-a^{2}}{2}\right)x+by \\
au+\left(\frac{a^{2}-b^{2}}{2}\right)v + \left(1+\frac{a^{2}-b^{2}}{2}\right)x-by \\
bv+vx+y
\end{array} \right).
\end{equation}
Using (\ref{E:125}) we finally get
\begin{equation}\label{E:149}
\xi'^{1}=\xi^{1}+a\hspace{1cm}\xi'^{2}=\xi^{2}+b.
\end{equation}
Hence $\Lambda(\textbf{a})$ generate the translation subgroup on
$\xi^{1}$ and $\xi^{2}$.

The special conformal transformations can be obtained in similar
way. Let $\textbf{c}=(c,d)$ be a two dimensional vector. Then
\begin{equation}\label{E:150}
\Lambda(\textbf{c})=e^{cC+dD}=\left( \begin{array}{cccc}
1 & -c & c & 0 \\
c & 1+\frac{d^{2}-c^{2}}{2} & \frac{c^{2}-d^{2}}{2} & -d \\
c & \frac{d^{2}-c^{2}}{2} & 1+\frac{c^{2}-d^{2}}{2} & -d \\
0 & -d & d & 1
\end{array} \right).
\end{equation}
Applying this transformation to $X^{T}=(u,v,x,y)$ we get
\begin{equation}\label{E:150}
\left( \begin{array}{c}
u' \\
v' \\
x' \\
y'
\end{array} \right)=\left( \begin{array}{cccc}
u-cv+cx \\
cu+\left(1+\frac{d^{2}-c^{2}}{2}\right)v + \left(\frac{c^{2}-d^{2}}{2}\right)x -dy\\
cu+\left(\frac{d^{2}-c^{2}}{2}\right)v + \left(1+\frac{c^{2}-d^{2}}{2}\right)x -dy\\
-dv+dx+y
\end{array} \right).
\end{equation}
Using again (\ref{E:125}) we get
\begin{equation}\label{E:151}
\xi'^{1}=\frac{\xi^{1}-c\left(\xi\cdot\xi\right)}{1-2\left(\xi\cdot\textbf{c}\right)+\left(\textbf{c}\cdot\textbf{c}\right)\left(\xi\cdot\xi\right)}\hspace{1cm}\xi'^{2}=\frac{\xi^{2}-d\left(\xi\cdot\xi\right)}{1-2\left(\xi\cdot\textbf{c}\right)+\left(\textbf{c}\cdot\textbf{c}\right)\left(\xi\cdot\xi\right)},
\end{equation}
where the inner product is with respect to the metric
$\textrm{diag}=(-1,1)$. Hence $\Lambda(\textbf{c})$ generate the
special conformal subgroup on $\xi^{1}$ and $\xi^{2}$.




Finally we will transform the coordinates $\xi^{1}$ and $\xi^{2}$
to complex coordinates in $\mathbb{C}^{2}$. We have
\begin{equation}\label{E:155d}
\xi^{1}+\xi^{2}=\tan\left(\frac{\lambda+\theta}{2}\right)\hspace{1cm}\xi^{1}-\xi^{2}=\tan\left(\frac{\lambda-\theta}{2}\right).
\end{equation}
If we define
\begin{equation}\label{E:155e}
z_{1}\equiv\frac{1+i(\xi^{1}+\xi^{2})}{1-i(\xi^{1}+\xi^{2})}\hspace{1cm}z_{2}\equiv\frac{1+i(\xi^{1}-\xi^{2})}{1-i(\xi^{1}-\xi^{2})},
\end{equation}
then
\begin{equation}\label{E:155f}
z_{1}=e^{i(\lambda-\theta)}\hspace{1cm}z_{2}=e^{i(\lambda+\theta)}.
\end{equation}
If now we make $\tau=i\lambda$ then
\begin{equation}\label{E:155f}
z_{1}=e^{\tau}e^{-i\theta}\hspace{1cm}z_{2}=e^{\tau}e^{i\theta}.
\end{equation}
We can consider $z_{1}$ and $z_{2}$ as complex conjugate of each
other and define on the complex plane. However following the usual
approach to Conformal Field Theory \cite{pdFranpMatdSene97} they
can be considered as independent complex variables and define
$\mathbb{C}^{2}$. At this point we could apply the standard
Conformal Field Theory to our problem by using $z_{1}$ and $z_{2}$
as our complex variables.

\end{document}